\documentclass[a4paper,amsmath,amssymb,pre,twocolumn,groupedaddress,superscriptaddress]{revtex4-2}

\usepackage{amsmath,systeme}
\usepackage[margin=1.5cm]{geometry}
\usepackage{graphicx}  
\usepackage{caption}
\usepackage{subcaption}

\usepackage{hyperref}
\usepackage{dcolumn}     
\usepackage{bm}    
\usepackage{color}  
\usepackage{array}
\usepackage{ragged2e}
\def\bd {\begin{displaymath}}
\def\ed {\end{displaymath}}

\begin{document}
\title{Infectious Disease Transmission In A Modified SEIRS model}
\author{Kasturi Banerjee}
\affiliation{Department of Physics, University of Calcutta, 
		Kolkata 700 009, India}
\affiliation{St. Augustine's Day School, Barrackpore, 64 Barrack Road, Kolkata 700 012, India}
\author{Subhankar Ray}
\affiliation{Department of Physics, Jadavpur University, 
	Calcutta 700 032, India}
\author{Jayalakshmi Shamanna}
\email[corresponding author: ]{jlsphy@caluniv.ac.in}
\affiliation{Department of Physics, University of Calcutta, 
	Kolkata 700 009, India}		

\begin{abstract}
Compartmental models like the Susceptible-Infected-Recovered (SIR)\cite{Kermack1927} and its extensions such as the Susceptible-Exposed-Infected-Recovered (SEIRS)\cite{Ottar2020,Ignazio2021,Grimm2021,Paoluzzi2021} are commonly used to model the spread of infectious diseases. We propose here, a modified SEIRS, namely, an SEIRSD model which comprises of (i) a reverse transmission from exposed to susceptible compartment to account for the probabilistic character of disease transmission seen in nature, and (ii) inclusion of mortality caused by infection in addition to death by other causes. We observed that, a reverse flow from exposed to susceptible class, has a significant impact on the height of infection peaks and their time of occurrence. In view of the recent surges of Covid-19 variants, this study is most relevant.
\end{abstract}

    \maketitle

    \section{Introduction}
    \label{intro}
    Transmission of disease is a complex dynamical process that varies with time and 
place across the world. It is controlled by several factors like nature of the pathogen 
and its physiology within the host body, social conditions of the diseased individuals etc. 
Breakout of COVID-19 pandemic in 2019 led to a renewed interest in epidemiology and 
the propagation of 
infectious diseases through populations. An important goal of epidemiological 
study is to
presage the extent and severity of the spread of infection among the population and 
assess whether the epidemic may escalate to a pandemic.
Epidemiological models, therefore, attempt to incorporate real-life scenarios by including imposition of preventive measures, public health advisories and effects of migration etc.    
    
 \section{SEIRS Model}
    In a recent article, Ottar et.al have proposed an SEIRS model with demographic
contribution\cite{Ottar2020}. An 
exposed class (E) has been added to SIR model, so that there is a time lag or 
latent period between getting infected and becoming infectious. Death due to 
natural causes is balanced by births, occurring at the same rate $\mu$. 
However, the rate of death due to infection, denoted by $\alpha$, 
is set to zero ($\alpha=0$).  
After recovery, immunity wanes over time and there is a flow into the susceptible compartment (the last S in SEIRS) at a rate $\omega$.  
The susceptible class remains populated and secondary infection peaks appear. 

 \section{Modified SEIRS : SEIRSD Model}
In this article, an SEIRSD model is proposed to extend and generalize the above SEIRS model. 
In addition to susceptible ($S$), exposed ($E$),  infected ($I$) and recovered ($R$), an infection induced deceased class ($D_i$) is considered, with total population $N = S + E +I +R+D_i $.
Two notable and distinct features of this SEIRSD model are as follows.
\begin{itemize}
	\item Disease transmission is probabilistic in nature. Not everyone
who is exposed to infection, gets infected. The non-infected people in the exposed group return to the susceptible group at a rate $\beta^{ \prime}$. This parameter, which represents the likelihood of not getting infected despite exposure to infectious population, is dependent on precautionary measures such as 
frequent sanitization, social distancing and wearing of masks.
This parameter lowers the infection peaks
and increases the interval between secondary peaks. This is a significant improvement over the SEIRS model where the parameters used in the model have no effect on
peak heights of infection and minor effect on interval between successive peaks.

As reflected in COVID-19 data\cite{Ourworldindata}, the peak heights and inter-peak 
intervals vary from country to country. The differing peak heights and inter-peak intervals, reflect to some extent, 
the level of enforcement, and the resultant efficacy of 
public health measures like social distancing, mandatory wearing of masks, screening,  
quarantine and lock-down which were imposed in several countries all over the world.

\item  Infection induced mortality class $D_i$, is considered, in addition to demographic outflow, $D_n$, i.e, death due to other causes, as, despite best efforts, death due to infection during the pandemic, was a grim reality that the world witnessed. 
\end{itemize}
\begin{figure}[h!]
\centering
\includegraphics[width=0.8\linewidth]{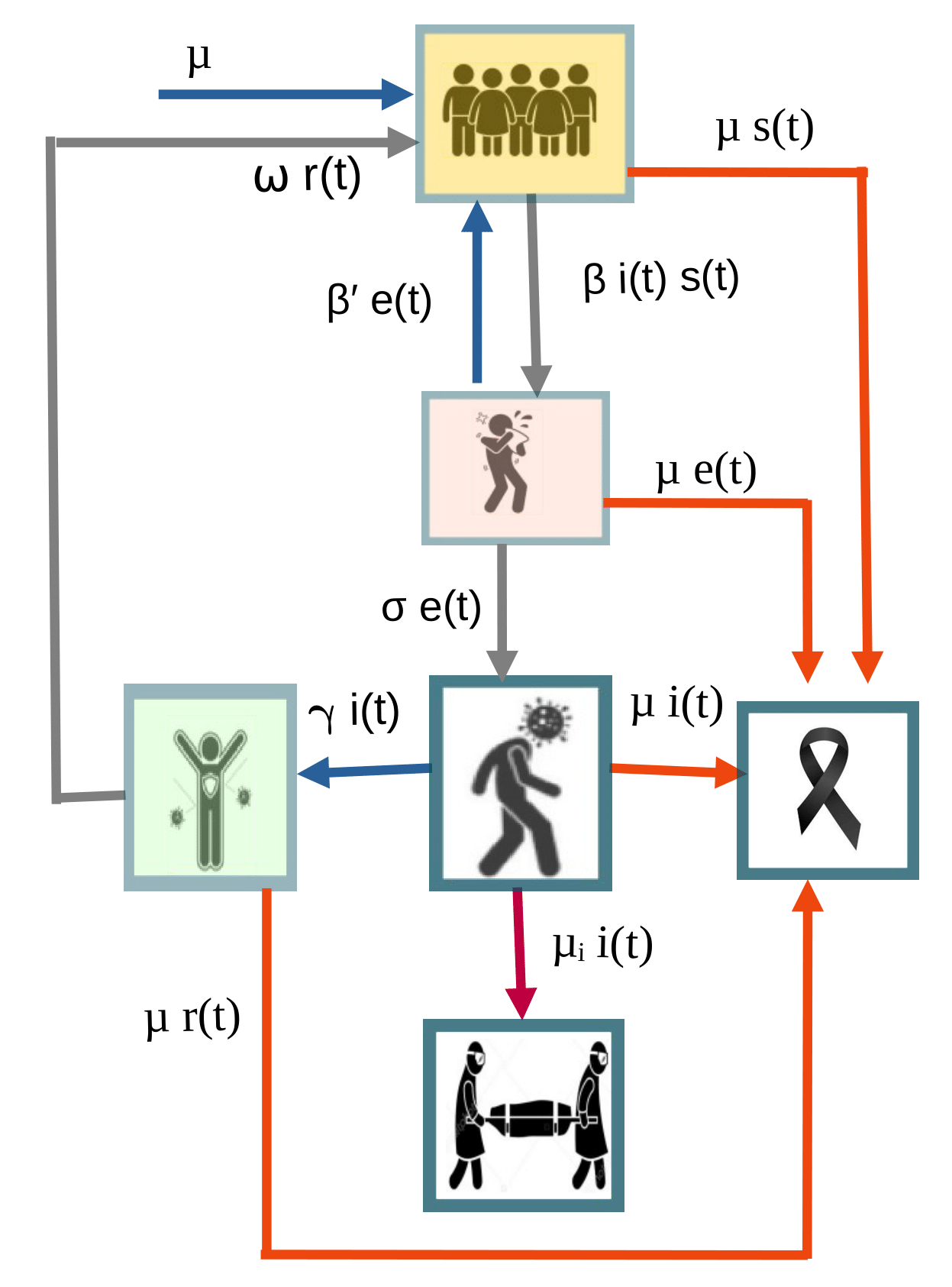}
\caption{Inflows and outflows in the SEIRSD model}
\label{Schematic}
\end{figure}

The dynamics of an epidemic are on a far shorter time scale, compared to the time for demographic flows, i.e., natural births and deaths. Hence, in our model, the demographic inflow and outflow rates, denoted by $\mu$, are taken to be the same,
and do not cause significant change in population size,
over the time scale of transmission of the disease.
The initial population is normalized as, 
\begin{equation}
	s(0) + e(0) + i(0) + r(0) + d_i(0)= 1.
\end{equation} 
The time derivatives of susceptible, exposed, infected, recovered and deceased (due to infection) classes, add up to zero, 
\bd
\frac{d}{dt}(s(t)+e(t)+i(t)+r(t)+d_i(t))=0
\ed
as an infected person, either recovers or succumbs to the disease.

Susceptible individuals move to exposed class after coming in contact with the infected class.
\begin{equation}
	\frac{ds}{dt} = \mu - \beta \ i(t)s(t) + \beta^{ \prime}\ e(t) + \omega \ r(t) - \mu \ s(t) \; ,
\end{equation}

The disease transmission rate of the infected class is taken as $\beta$. It depends on the nature of pathogens, contact rates and infectiousness of the infected class.

For the exposed class, the only inflow is from the susceptible class.
\begin{equation}
	\frac{de}{dt} = \beta \ i(t) s(t)-\sigma \ e(t)-\beta^{'}\ e(t)-\mu \ e(t)
\end{equation}	  

Exposed individuals who get infected, move to the infected class at a rate $\sigma$. The time lag between getting exposed and becoming infected, specifically, the onset of symptoms for the symptomatic individuals, is $\sigma^{-1}$.
Those who do not get infected, return to the susceptible class at a rate $\beta^{ \prime}$. 

Infected individuals recover 
at rate $\gamma$, or in unfortunate cases, succumb to the disease at a rate $\mu_{i}$ as indicated in Fig\ref{Schematic}.

\begin{equation}
	\frac{di}{dt} = \sigma \ e(t) - \gamma \ i(t) - (\mu_{i} + \mu) \ i(t)    	
\end{equation}   

The recovery rate $\gamma$ is indicative of how soon an infected individual recovers and moves to the recovered class.

\begin{equation}
	\frac{dr}{dt} = \gamma \ i(t)-\omega \ r(t)-\mu \ r(t)
\end{equation}

Post recovery, there is loss of immunity over an average time  $1/\omega$, whence the recovered individuals become susceptible to re-infection.

From each compartment, there is demographic outflow, due to natural causes. 
The rate of mortality caused by infection, will vary according to the severity of infection, viral load, pre-existing medical and physiological conditions of the infected individuals and timely access to medical facility as well as the quality of medical care etc. 
\begin{equation}
	\frac{d(d_n)}{dt} = \mu \ (s(t)+e(t)+i(t)+r(t))
\end{equation}
\begin{equation}
	\begin{split}
		\frac{d(d_i)}{dt} = \mu_{i} \ i(t)
	\end{split}
\end{equation}


\section{Results and Observations}    
The peak heights of infection, inter-peak intervals and endemic equilibrium
are important indicators of the severity of infection in a population. They serve as important inputs for outbreak response by the healthcare system in a country.
In the SEIRS model by Ottar et al\cite{Ottar2020}, two parameters, $\sigma$, related to latency and $\omega$, related to duration of immunity,  are considered.
Varying these parameters, the inter epidemic interval $T_{E}$, and endemic equilibrium can be controlled. No change however, is reported in heights of infection peaks.

However, as stated earlier, the peak heights and inter-peak intervals differ from country to country. 
For a particular disease, such as SARS-CoV-2 coronavirus, $\sigma$ is nearly constant in a country. Hence, attempting to control infection peaks by changing $\sigma$ is not very realistic.
In the present SEIRSD model, an additional parameter $\beta^{\prime}$ for the backflow from exposed to susceptible
is introduced, This allows for better control over infection peak heights, as well as inter-peak interval and
endemic equilibrium. This is consistent with the different peak heights and inter-peak intervals as per Covid-19 data of
different countries\cite{Ourworldindata}.

 \subsection{Susceptible, Exposed and Infected classes}
\begin{figure}[h!]
		\centering
    	\includegraphics[width=0.8\linewidth]{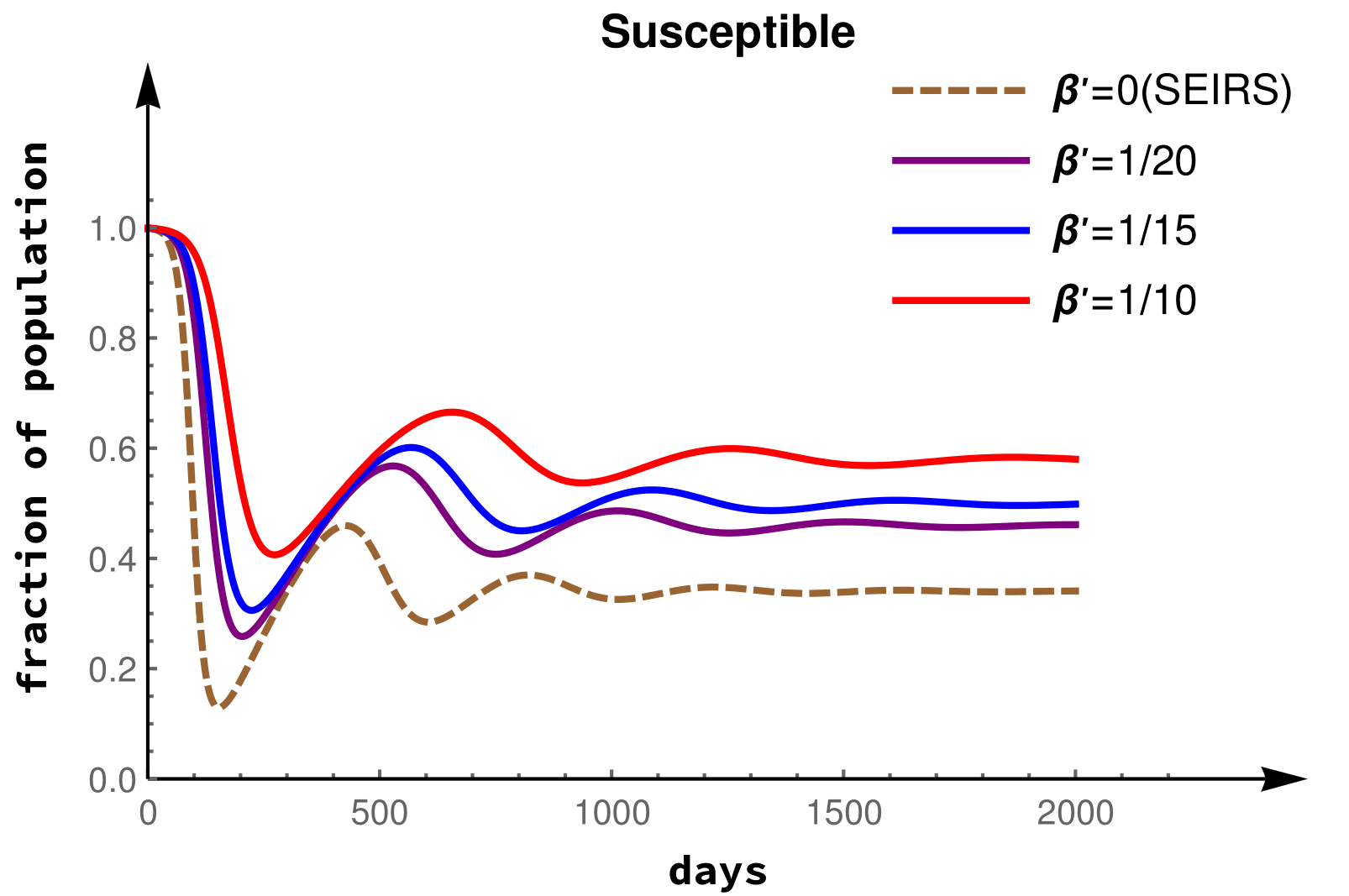}
    	\caption{Relative size of susceptible class for different $\beta^{\prime}$}
    	\label{SuscepE2S}
    \end{figure}

\begin{figure}[h!]
		\centering
		\includegraphics[width=0.8\linewidth]{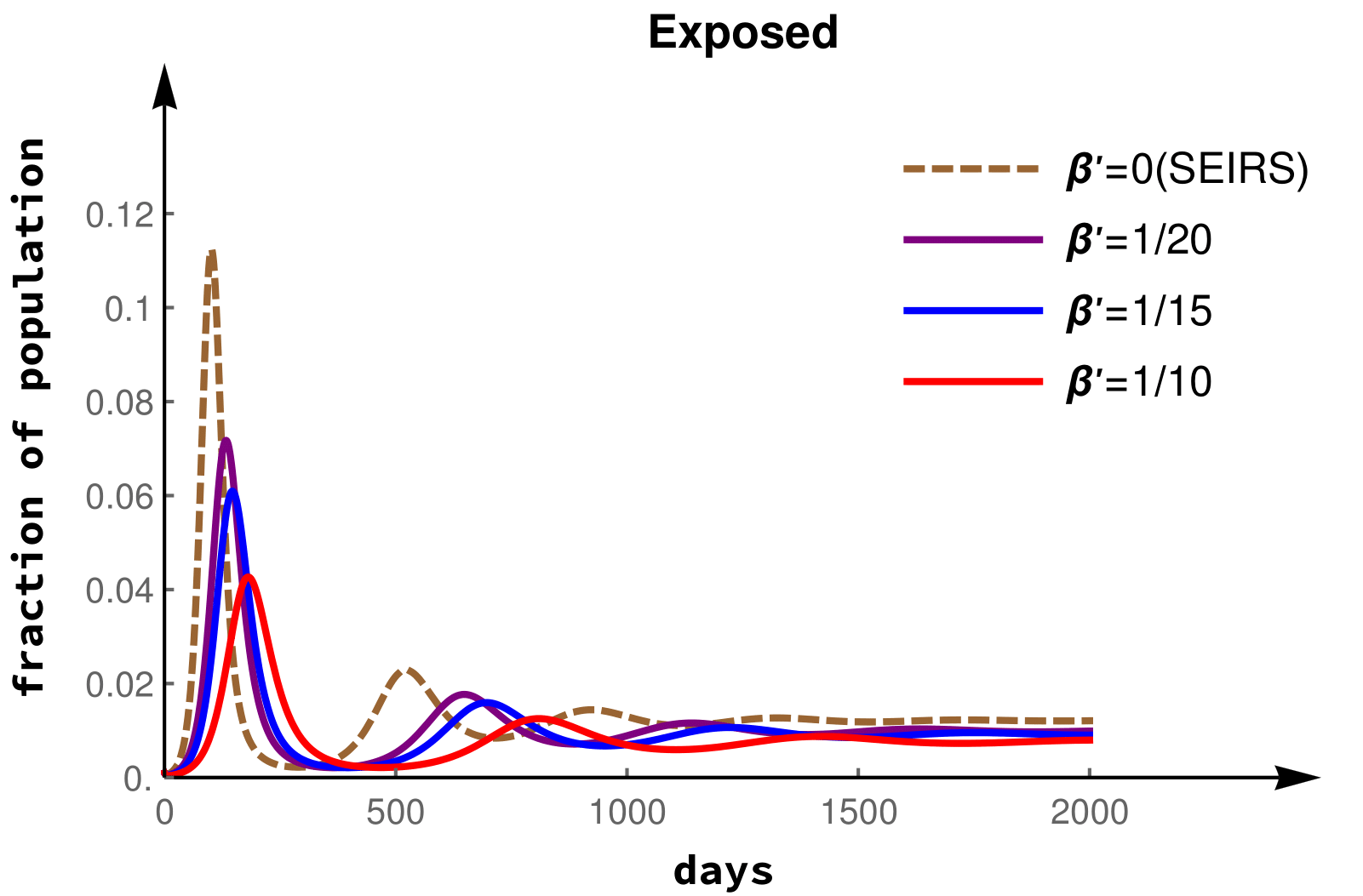}
		\caption{Evolution of exposed class for different $\beta^{\prime}$}
		\label{ExpE2S}
	\end{figure}

\begin{figure}[h!]
		\centering
		\includegraphics[width=0.8\linewidth]{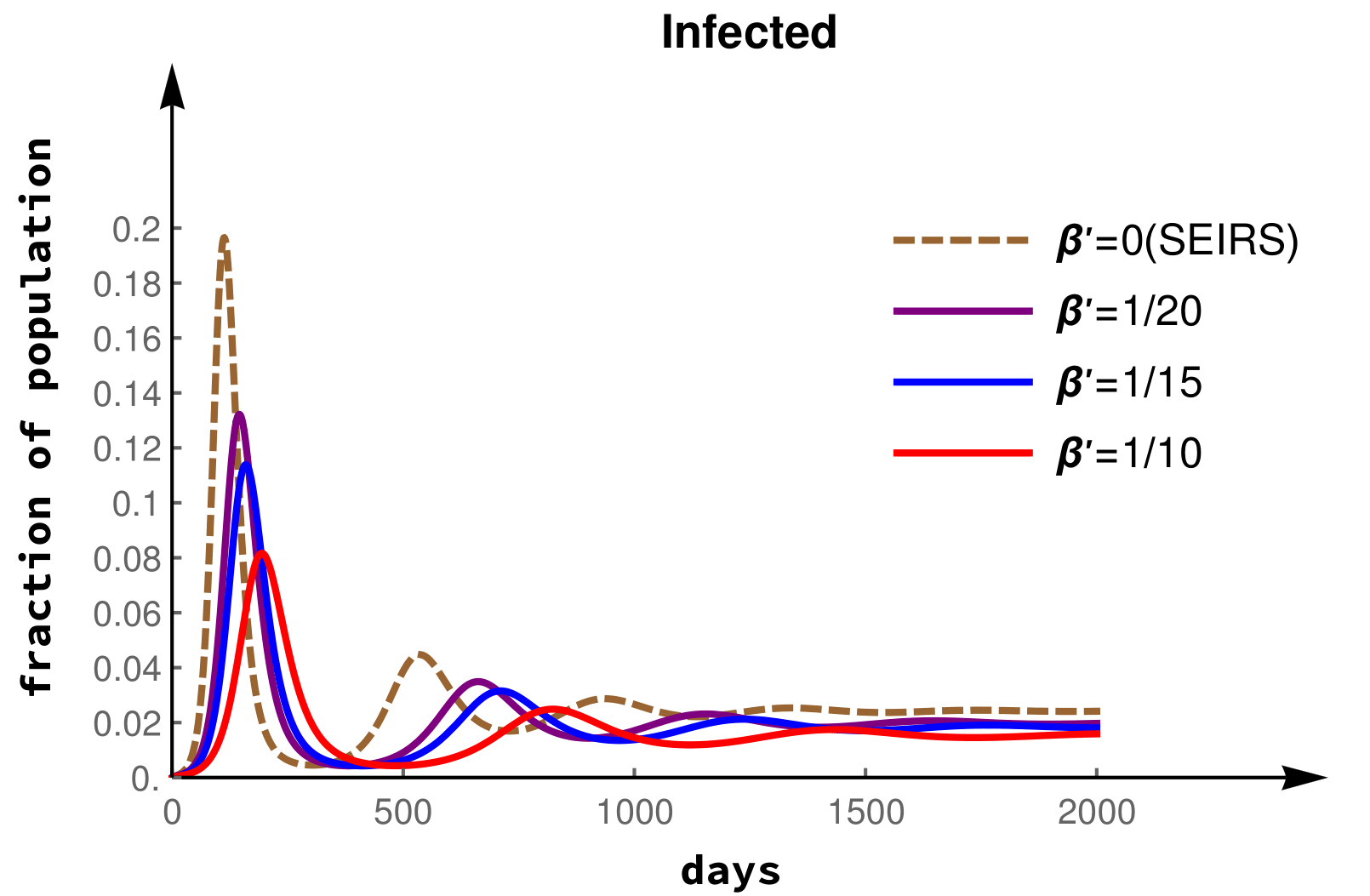}
		\caption{Primary and secondary peaks of infection for different $\beta^{\prime}$}
		\label{InfE2S}
	\end{figure}

 The effect of reverse flow from exposed to susceptible class at a rate $\beta^{\prime}$ is examined. 
The evolution of the susceptible class, the exposed class and the infected class is shown in Fig:\ref{SuscepE2S}, Fig:\ref{ExpE2S} and Fig:\ref{InfE2S}. For comparison, corresponding plots for the SEIRS model, where there is no exposed-to-susceptible backflow,  is also shown.

It is observed that,
\begin{itemize}
	\itemsep=-1.0pt 
	\item The susceptible class evolves at slower rates as $\beta^{\prime}$ is increased.
	\item Peak heights in the exposed and infected classes are lower and peaks appear at later times. 
	\item The number of secondary peaks decreases, eventually approaching endemic equilibrium with lower values.
\end{itemize}

In addition, the fluctuations in the susceptible class damp out sooner (Fig:\ref{SuscepE2S}) when $\beta^{\prime}$ is greater than $\sigma$, the  rate of flow from exposed to infected class. The secondary waves of infection take longer time to peak, providing healthcare administration and medical fraternity much needed time to take preparatory measures.
Thus safety measures like mandatory wearing of masks, proper hygiene, social distancing, all contribute to increasing $\beta^{\prime}$ and 
can go a long way in keeping the spread of infection within manageable limits.

Using parameter values $1/\sigma$ = 7 days, $1/\omega$ = 1 year and $1/\beta^{\prime}$ = 10 days, the first peak height drops to 0.08 from 0.20 and the inter-peak interval extends to 1.72 years. Endemic equilibrium which is 2.4\% in SEIRS model decreases to 1.6 $\%$ (Table:\ref{betaprime}) in SEIRSD model.
\begin{table} [h!]
{\renewcommand{\arraystretch}{1.4}%
\begin{tabular}{|c|c|c|c|}
	\hline
	$\beta^{\prime}$ & Position & Height &Endemic \\
	days$^{-1}$ & (days) &  & equilibrium (\%) \\
	\hline
	1/30 & 132.9 & 0.1528 & 2.20 \\
	\hline
	1/25 & 137.5 & 0.1452 & 2.18 \\
	\hline
	1/20 & 145.4 & 0.1334 & 2.05 \\
	\hline
	1/18 & 149.1 & 0.1266 & 2.00 \\
	\hline
	1/15 & 158.8 & 0.1146 & 1.90 \\
	\hline
	1/12 & 175 & 0.0978 & 1.74 \\
	\hline
	1/10 & 193.4 & 0.0824 & 1.59 \\
	\hline
\end{tabular}}
\caption{Position of the first infection peak, peak height and endemic equilibrium for different $\beta^{\prime}$. Other parameters : $\mu = 76 \ $ years, $\beta = 0.21$, $1/\sigma = 7 \ $ days, $1/\gamma = 14 \ $ days, $\mu_{i} = 0.0$, $1/\omega = 1 \ $ year}
\label{betaprime} 
\end{table} 
 
\begin{figure}[h!]
	\centering
	\includegraphics[width=0.7\linewidth]{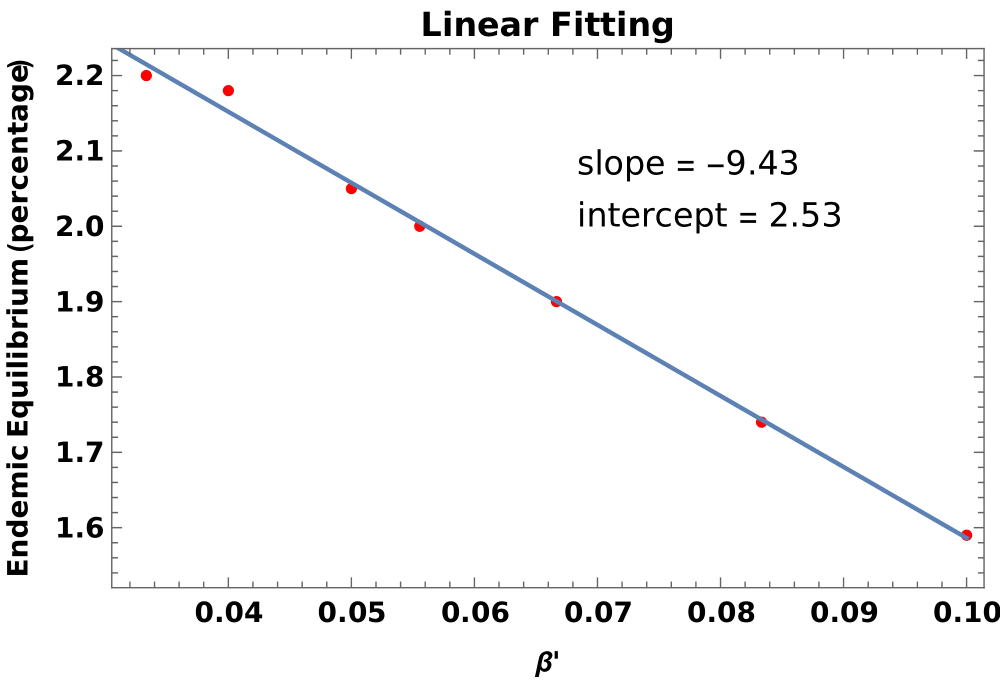}
	\caption{With increase in $\beta^{\prime}$ endemic equilibrium (in \%) decreases linearly.}
	\label{data2_lf}
\end{figure}

Interestingly, the endemic equilibrium has a linear relation with $\beta^{\prime}$, as is shown in Fig:\ref{data2_lf}.

\subsection{Recovery from infection and mortality caused by infection}
	\begin{figure}[hbt!]
		\centering
		\includegraphics[width=0.8\linewidth]{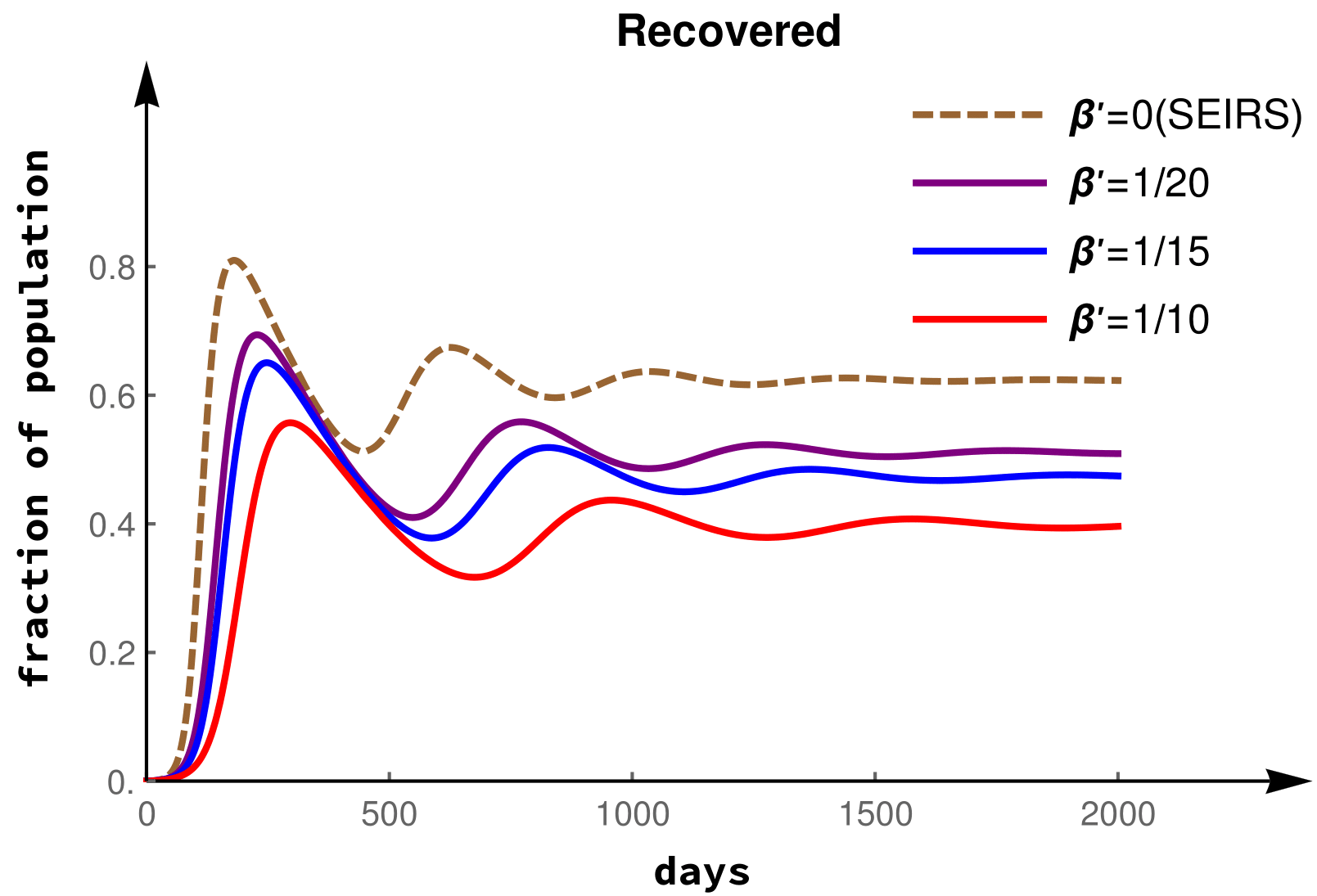}
		\caption{Relative size of recovered class for different $\beta^{\prime}$. Death due to infection, lowers the recovered class peaks further.}
		\label{RecvE2S}
	\end{figure}

As noted earlier, the backflow $E \to S$ and infection induced mortality, both reduce the size of the infected class when compared with the SEIRS model. In consequence, the recovered class is also smaller.

During an epidemic or pandemic, mortality caused by infection is a sad reality. 
It is known that mortality rates of infection for different variants of SARS-Cov2 are not the same. While the $\alpha$- variant caused high mortality  in Europe, the $\delta$- variant caused widespread infection and death in India and other Southeast Asian countries as per data published by (WHO) World Health Organization \cite{WHO}.

\begin{figure}[h!]
	\centering
	\includegraphics[width=0.75\linewidth]{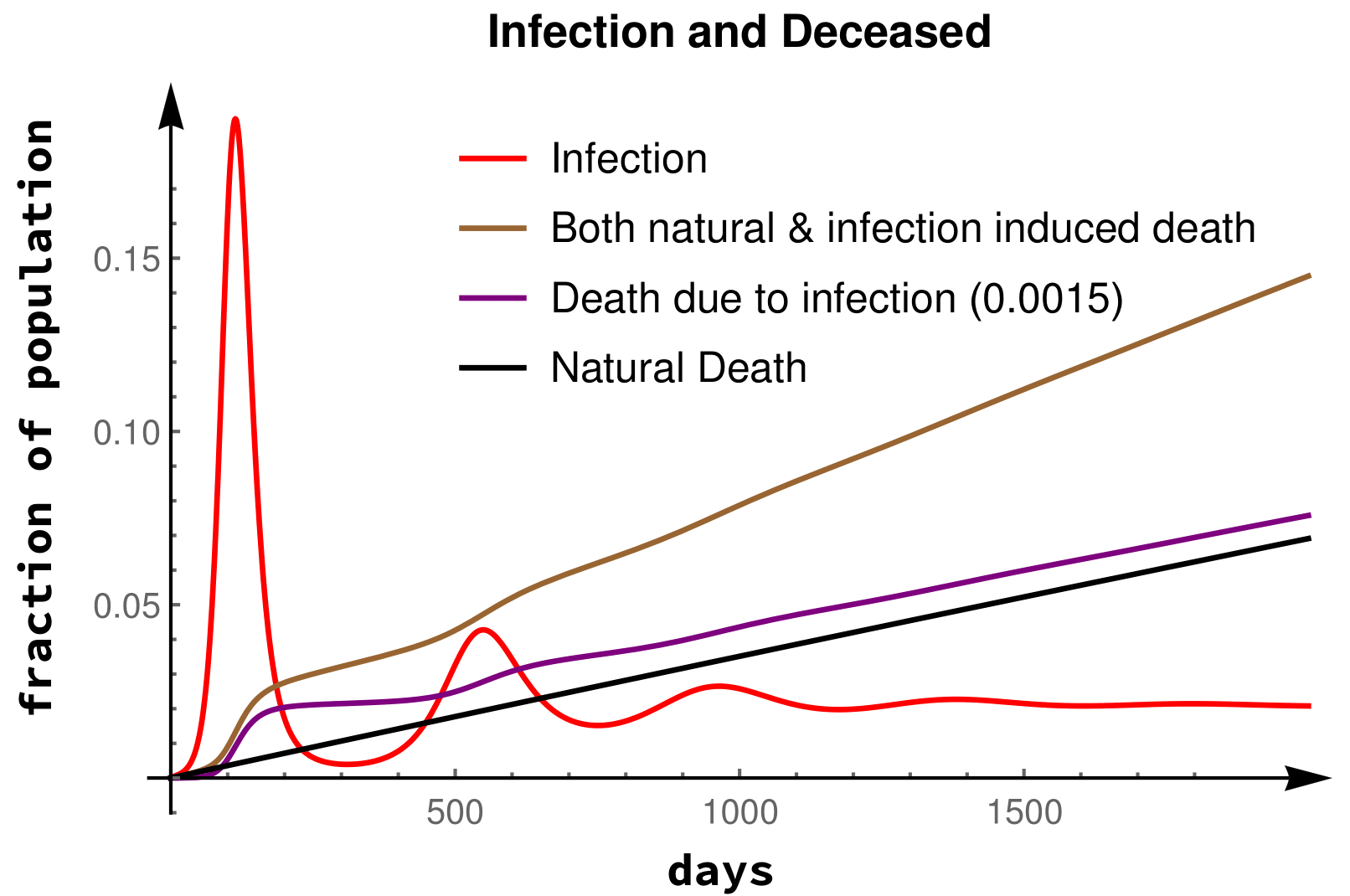}
	\caption{Variation in relative sizes of Infected and Deceased classes.}
	\label{Inf_Death}
\end{figure}

The SEIRSD model which considers mortality due to infection, in addition to natural deaths, can provide broad guidance towards bolstering public health infrastructure and implementation of effective healthcare measures to contain the infection and reduce
mortality due to infection. 
\begin{figure}[h!]
	\centering
	\includegraphics[width=0.75\linewidth]{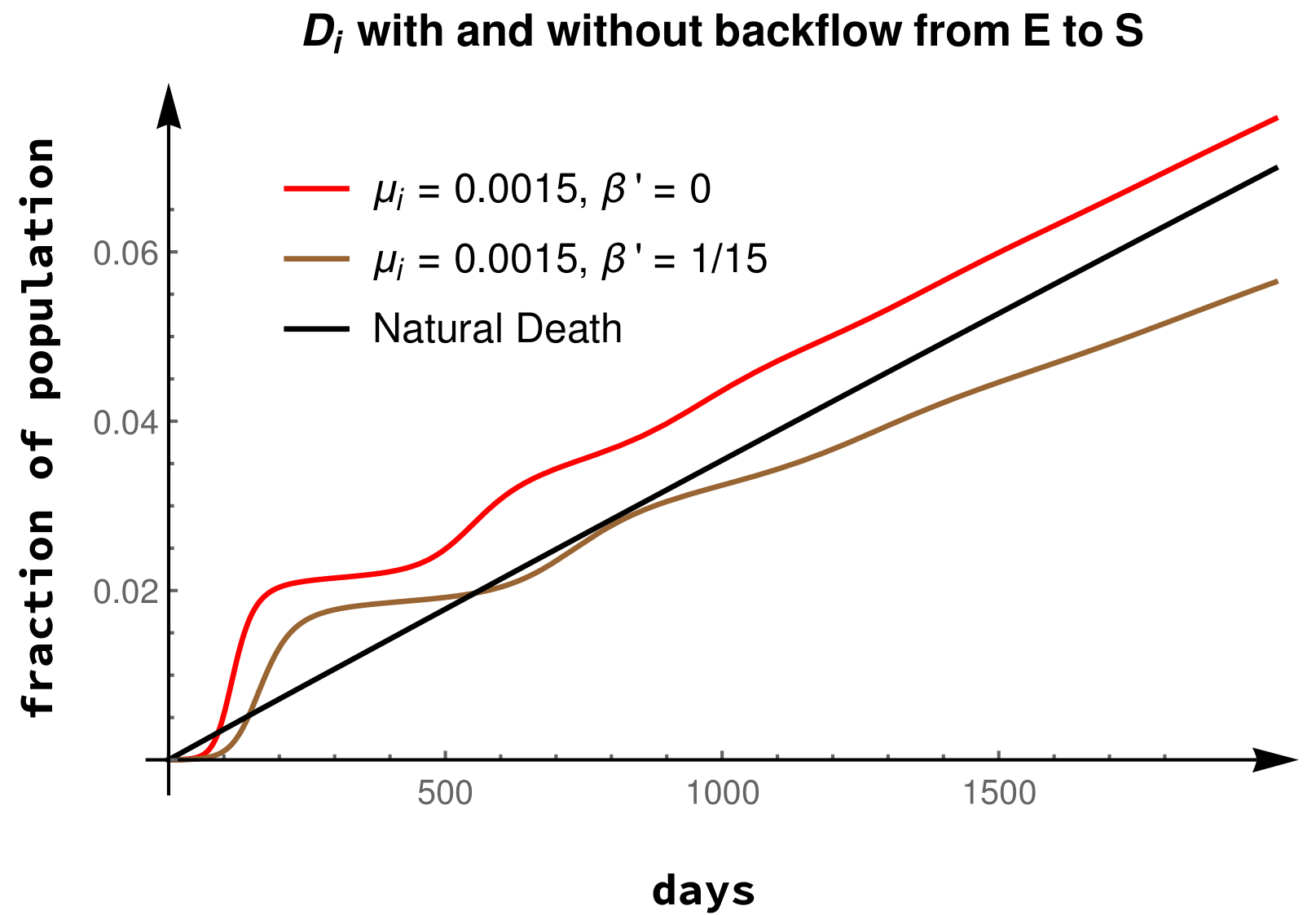}
	\caption{Death due to infection with and without backflow from exposed to susceptible}
	\label{di_e2s}
\end{figure}
It is observed that (Fig:\ref{Inf_Death}), the infection and mortality caused by infection, peak at 110 and 159 days respectively, implying infection induced death happens approximately 49 days after the infection has reached its peak. This is consistent with WHO reports\cite{WHO2}.
Total deaths (deaths due to both natural and infection related causes) and deaths caused by infection, are also shown in Figure \ref{Inf_Death}. Initial fluctuations are due to the wave-like rise and fall of infection and are followed by a linear increase, reflecting the existence of a constant fraction of infection present in the population. Introducing backflow $E \to S$, significantly lowers mortality due to infection as depicted in Figure \ref{di_e2s}.
  
\section{Conclusion}
Multi-compartmental models give an improved visualisation of the complex process
of disease transmission. As a realistic approach, we have presented an SEIRSD model, where (i) a fraction of the exposed population reverts to susceptible, instead of developing infection and
(ii) infected population show an increased mortality caused mostly due to infection.
As the rate of reverse transmission parameter, $\beta^{\prime}$, increases, epidemic shows gradual decrease in infection peaks. In the long run, a small percentage of the total population remain infected, indicating endemic equilibrium.
To contain the disease and reduce mortality, strict public health measures such as, mandatory wearing of masks and social distancing may be implemented. At a more organized level, measures such as quarantine and hospitalization  can contribute significantly towards containing the spread of infection. 
As an extension and continuation of our model, we propose to introduce in a future work, multiple subcategories having different infectiousness and mortality rates. This will be a step forward in progressively improving such models to incorporate real life scenarios and hence better represent disease transmission in nature.

\end{document}